\documentclass[lettersize,journal]{IEEEtran}
\usepackage{amsmath,amsfonts}
\usepackage{algorithmic}
\usepackage{epstopdf}
\usepackage{epsfig,amssymb,amsbsy,verbatim,array}
\usepackage{algorithm}
\usepackage{array}
\usepackage[caption=false,font=normalsize,labelfont=sf,textfont=sf]{subfig}
\usepackage{textcomp}
\usepackage{stfloats}
\usepackage{hyperref}
\usepackage{url}
\usepackage{verbatim}
\usepackage{graphicx}
\usepackage{cite}
\usepackage{multirow}
\usepackage{multicol}
\usepackage{bm}
\usepackage{verbatim}
\usepackage{color}

\begin{document}

\title{Channel Modeling Aided Dataset Generation for AI-Enabled CSI Feedback:  Advances, Challenges, and Solutions}

\author{ ~\IEEEmembership{Yupeng Li, Gang Li, Zirui Wen, Shuangfeng Han, Senior Member, IEEE, Shijian Gao, Member, IEEE,
Guangyi Liu, Member, IEEE, and Jiangzhou Wang, Fellow, IEEE}
\thanks{Corresponding author: (liyupengtx@126.com; liyupeng@chinamobile.com) }%
}

\markboth{IEEE ,~Vol.~14, No.~8, August~2023}%
{Shell \MakeLowercase{\textit{et al.}}:  }

\IEEEpubid{}

 \maketitle

\begin{abstract}
The AI-enabled autoencoder has demonstrated great potential in channel state information (CSI) feedback in frequency division duplex (FDD) multiple input multiple output (MIMO) systems. However, this method completely changes the existing feedback strategies, making it impractical to deploy in recent years. To address this issue, this paper proposes a channel modeling aided data augmentation method based on a limited number of field channel data. Specifically, the user equipment (UE) extracts the primary stochastic parameters of the field channel data and transmits them to the base station (BS). The BS then updates the typical TR 38.901 model parameters with the extracted parameters. In this way, the updated channel model is used to generate the dataset. This strategy comprehensively considers the dataset collection, model generalization, model monitoring, and so on. Simulations verify that our proposed strategy can significantly improve performance compared to the benchmarks.
\end{abstract}


\section{Introduction}
\IEEEPARstart{I}{n} massive multi-input multi-output (MIMO) systems,  it is crucial for transceivers to obtain accurate channel state information (CSI) to enhance spectrum efficiency and capacity density. However, in the frequency division duplex (FDD) system, the downlink and uplink channels exhibit different channel fading characteristics due to their operation at different frequencies. Consequently, the user equipment (UE) must provide the base station (BS) with precise and comprehensive feedback of the downlink CSI through the uplink channel. One challenge of CSI feedback is the significant uplink resource overhead, particularly in massive MIMO systems.

To address this issue, various solutions have been proposed in the 3rd Generation Partnership Project (3GPP). For example, the eType \uppercase\expandafter{\romannumeral2} codebook \cite{codebook} compresses the downlink CSI based on sparsity in the spatial and frequency domains. Additionally, to further enhance recovery accuracy and reduce feedback overhead, a deep learning (DL) based autoencoder named CsiNet has been proposed \cite{csinet}. Subsequently, various works have been proposed to adapt to more scenarios \cite{2} or to address the dependency of field data \cite{meta}. However, despite the autoencoder demonstrating superior performance over traditional codebook-based (e.g., eType II codebook) schemes, there are still critical challenges for its deployment in practical communication systems.
Firstly, the AI model requires extensive training with a vast channel dataset, resulting in significant data collection and air interface expenses. This becomes particularly challenging in dynamic communication environments, where the channel dataset needs frequent updates, making deployment of the autoencoder difficult. Moreover, the generalization of the AI model is another crucial issue. As the UE moves to another scenario, the corresponding channel distribution properties may differ significantly.
Finally, training the AI model is a time-consuming task, and there are no clear conclusions on how to construct "standard" offline AI models provided for the UE to choose or switch, suitable for changing communication environments. The aforementioned issues are closely related to the entire life cycle management (LCM) of the AI model, which includes data collection, AI model training, model inference, model generalization, model monitoring, model selection, and so on.

As one of the most crucial aspects, the data generation technique not only influences the generalization of the AI model but also determines the expense of the data collection. Data augmentation is an efficient tool for constructing diverse datasets and mitigating the overfitting problem, and it has been widely utilized in the fields of computer vision (CV) and natural language processing (NLP) \cite{noise}. However, traditional data augmentation methods are designed based on the characteristics of image or voice data, which are fundamentally different from channel data. Specifically, channel data is high dimensional in the receiving-transmitting space domain, time domain, delay domain, and so on. The data in different domains exhibit different distribution characteristics. Additionally, channel data has different sparse characteristics compared to image or voice data, stemming from the unique wireless propagation characteristics. These differences mean that traditional data augmentation methods may not work effectively with channel data. Furthermore, the time taken to generate the dataset and train the AI model is significant, especially in communication systems with high real-time requirements.

\begin{figure*}[!t]
\centering
\includegraphics[width=7.5in]{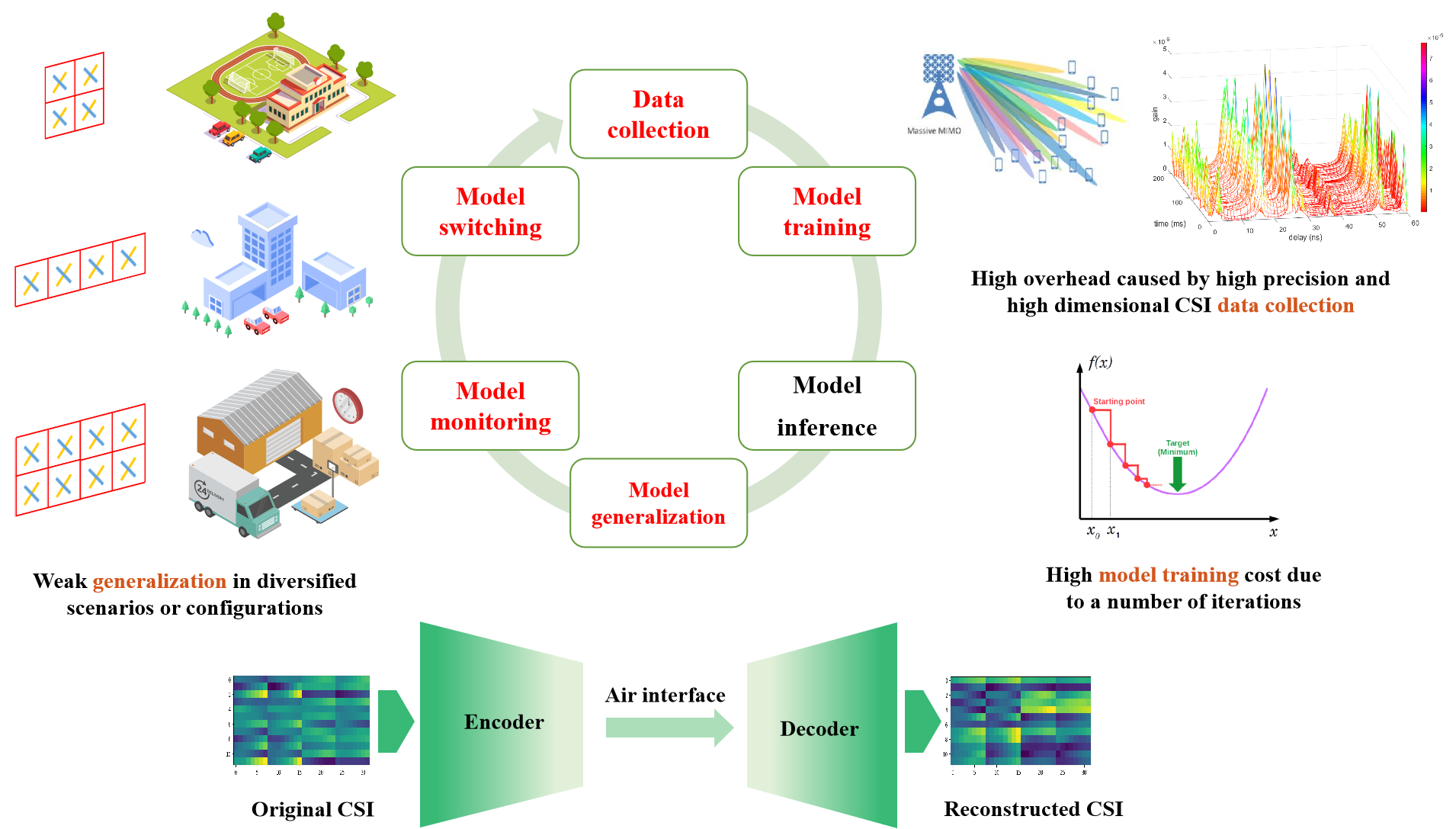}
\caption{The LCM for the two sided AI model.} \label{stand}
\end{figure*}

Motivated by the aforementioned requirements, a channel modeling enabled data generation method (CMDG) is proposed. This method focuses on optimizing data collection and augmentation, aiming to reduce data collection overhead and enhance model generalization. Specifically, field measurements are conducted to obtain downlink channel data. Subsequently, the primary channel parameters are extracted and used to analyze the stochastic parameters of the channel. As a result, the BS updates the stored typical TR 38.901 channel model parameters with the stochastic parameters, referred to as the scene-specific channel model (SSCM) hereafter. Secondly, a strategy for dataset construction is proposed to reduce the time taken to generate the dataset and train the model. The channel statistical parameters extracted from the field scenario can be utilized to distinguish the scattering environment and can be leveraged for dataset construction.
The main contributions of our work are as follows.
\begin{itemize}
\item We introduce a novel channel modeling enabled data augmentation method for CSI feedback, requiring only a limited amount of field data. This approach significantly reduces field measurement overhead and air interface expenses.
\item We propose a dataset construction strategy leveraging channel statistical parameters, effectively reducing data generation and training delays for the AI model.
\item Our work comprehensively addresses the issues of the AI model LCM, encompassing aspects such as data collection, air interface overhead, model generalization, model monitoring, and model switching.
\end{itemize}

The reminder of the paper is organized as follows. The standardization challenges are discussed in Section II. The channel modeling aided dataset augmentation is given in Section III. Simulation results are presented to verify the efficiency of our proposed strategy in enhancing the system performance in Section IV. Followed by the standardization impact in Section V. Finally, the conclusions and future work are illustrated in Section VI.

\section{Standardization Challenges}
The study item \cite{si} has been initiated in 3GPP RAN1 to investigate CSI compression and feedback using a two-sided autoencoder model. In this model, the original CSI is compressed to a quantized bitstream with an encoder model at the UE and transmitted to the BS via the air interface, where it is then uncompressed to a recovered CSI using a decoder model \cite{guo}. However, this two-sided model scheme may lead to several issues, such as a significant overhead in training data collection, data uploading overhead via the air interface, weak generalization, AI model inter-operability issues during the training phase, and more. Figure \ref{stand} illustrates the lifecycle management (LCM) for the AI model. Therefore, this section aims to comprehensively analyze these challenges in depth.

\subsection{Efficient Dataset Generation }
The simulation dataset often fails to encompass the full range of diverse radio channel environments, and there is typically a significant distribution gap between simulation and field channel data. Therefore, it is practical to use field datasets to train the AI model.

However, field measurements result in extensive data collection and air interface expenses, making it impractical to carry out. Training the AI model requires high-precision data; for instance, the size of a 32-antenna port CSI sample is approximately 3.3 kilobytes with a precision of float 32. Consequently, the total storage occupation is nearly 2 gigabytes for a 2-rank, 30 thousand sample dataset \cite{huawei}. If multiple antenna configurations or scenarios are considered, the dataset size multiplies, posing a substantial storage burden on the UE. Furthermore, the collected channel data acquired by the UE must be uploaded to the network for training, resulting in a significant overhead on the air interface.

Therefore, both industry and academia lack comprehensive consideration of the following questions related to data generation:
\begin{itemize}
    \item What type and precision of channel data should be collected by the UE? Should it be high-precision channel samples, quantized channel samples, statistical characteristics of the channel samples, or a combination of these?
    \item How can plentiful propagation characteristics of a scenario be collected to improve the generalization of the dataset (AI model) with limited samples?
    \item How many samples are sufficient to effectively train the AI model in a specific scenario? If limited samples are adequate, what is the efficient data augmentation method with a limited number of samples?
\end{itemize}

\subsection{Dataset Constructions Towards Generalization}
The performance of an AI model trained in one scenario (configuration) may degrade significantly when applied to another scenario (configuration), which is commonly referred to as a generalization issue. In wireless communication systems, configurations are flexible and based on communication requirements, primarily including communication bandwidth, port number, feedback bits, and other factors. If the configuration changes, the performance of a legacy AI model may degrade by approximately 3\% for bandwidth, over 20\% for port number, and 2\% for feedback bits, separately \cite{Nokia}. Channel scenarios vary, mainly including Urban Micro (UMi), Urban Macro (UMa), Indoor Hotspot (InH), and others. Experiments have shown that when a UE moves from UMi to InH, the performance degradation can be as high as 20\% \cite{Nokia}.

Therefore, there is a need to address how to construct the dataset to enhance the generalization of the AI model. Is it necessary to construct a "standardized" dataset consisting of multiple scenarios provided to the UE for model selection? Additionally, how can the training and model switching delay be minimized once the UE moves to a new scenario? These aforementioned issues are all open challenges that need to be addressed.

\section{Channel modeling aided dataset augmentation}

\subsection{Wireless Channel Feature Extraction}\label{modeling}

In this section, we generally describe some extracting methods of the channel parameters from the receiving frequency domain channel matrix $H_{f}$.

\subsubsection{Delay spread}
The time domain channel matrix $H_{t}$ can be derived by performing an inverse Discrete Fourier transform (IDFT) across the frequency dimension of the frequency domain channel matrix $H_{f}$. Utilizing the time domain channel matrix, we can compute the power delay spectrum, which allows us to determine the multipath delays and their associated power. Subsequently, the delay spread (DS) can be calculated by obtaining the mean value of the multipath delays, weighted by the corresponding multipath power \cite{extrpara}. The DS serves as a characterization of the multipath component (MPC) richness within the channels.

\subsubsection{Angle spread}
The frequency domain channel matrix $H_{f}$ is in the port domain, and therefore, it needs to be transformed into the angle domain channel matrix $H_{ang}$ using a two-dimensional Discrete Fourier transform (DFT). This transformation allows for the retrieval of receiving and transmitting angle values within sparse spatial directions. The angle spread (AS) value can then be computed by evaluating the angle values weighted by their corresponding gains \cite{901}. The AS serves as an effective indicator of the dispersion characteristics of multipaths.

\subsubsection{Ricean K factor}
To calculate the Ricean K factor (KF) in a wideband system, the distinguishable multipaths are initially extracted from the channel matrix. Subsequently, the KF can be derived from the ratio of the power of the deterministic MPC and the power of all the other stochastic MPCs \cite{extrpara}.

In this paper, we have provided a preliminary description of some methods for extracting multipath statistic parameters. For additional channel statistic parameters, such as cluster number, cluster DS, and cluster AS, readers can refer to \cite{ruisi}.


\subsection{Data Augmentation via Channel Modeling}
Data augmentation is a powerful technique for creating diverse and plentiful training datasets. It has been extensively utilized in the fields of CV and NLP \cite{noise}, involving methods such as noise injection, random erasing, and flipping, which are aimed at enhancing dataset diversity and alleviating overfitting issues.

While these methods are well-suited for traditional DL problems, they are designed based on the characteristics of image or voice data. However, channel data in communications exhibits significantly different characteristics compared to traditional DL problems. Specifically, channel data is high-dimensional, encompassing the receiving port domain, transmitting port domain, time domain, delay domain, and more. Data from different domains possess distinct distribution characteristics. The unique attributes of channel data render traditional data augmentation techniques ineffective in capturing the intrinsic characteristics of the channel data, thereby leading to poor performance in the wireless communication domain.

\begin{figure*}[!t]
\centering
\includegraphics[width=7.5in]{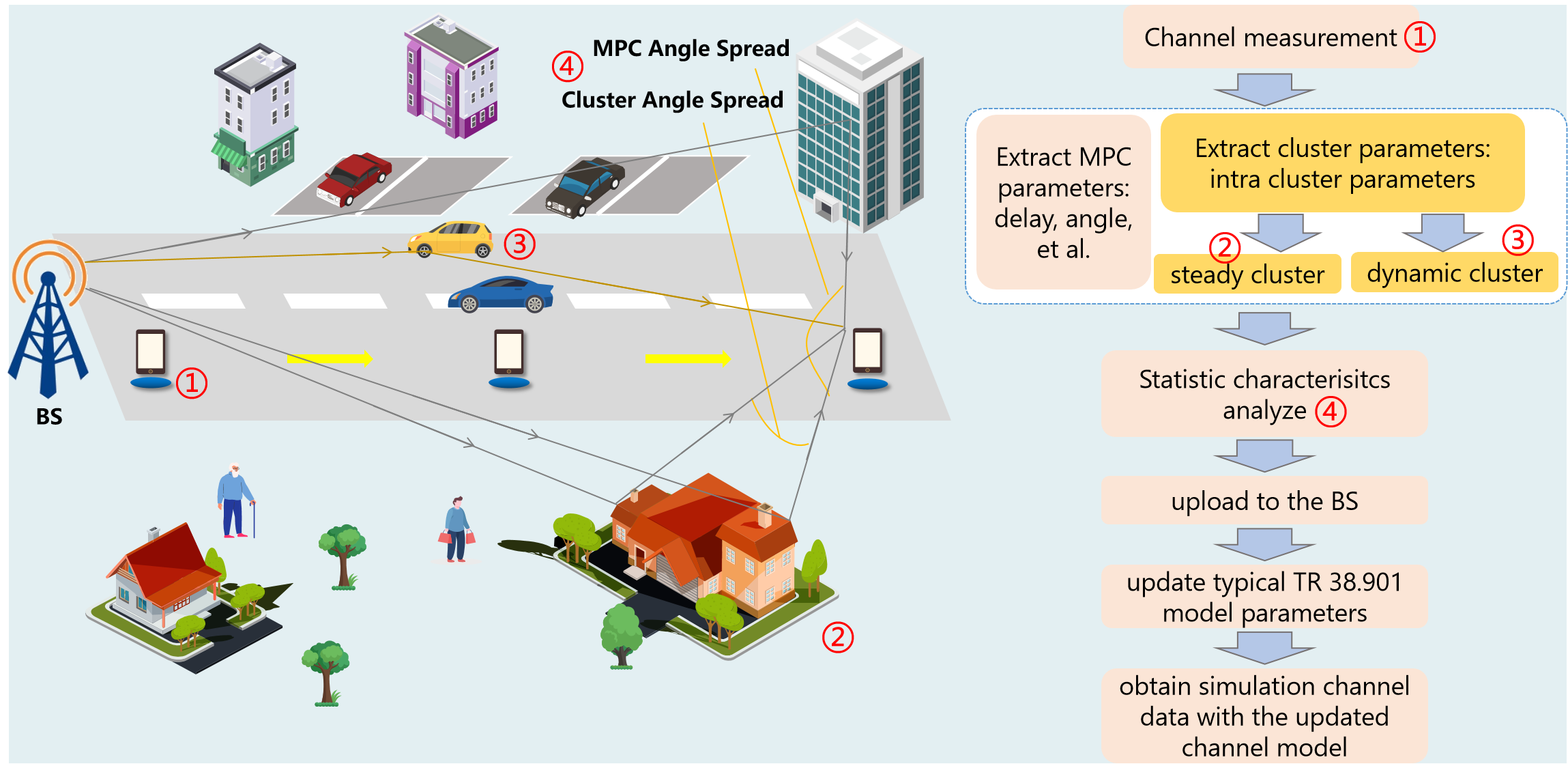}
\caption{Channel modeling based data augmentation method.}\label{flowpath}
\end{figure*}

The 3GPP channel model, TR 38.901 \cite{901}, represents the state-of-the-art standard for obtaining channel simulation data and effectively evaluating physical layer techniques. This model falls under the category of geometry-based stochastic channel models (GBSM) \cite{bigdata}, which captures the stochastic distribution characteristics of MPCs, such as DS, AS, and more. The statistical parameters within TR 38.901 are typically weighted or averaged by various companies, making it a widely used benchmark model in general scenarios. It is important to note that in specific locations, the channel parameters in TR 38.901 may not be entirely accurate. However, in the case of CSI feedback, the training dataset is expected to accurately reflect the actual channel propagation environment as much as possible.


Inspired by the GBSM channel modeling method, if one desires a large volume of simulation channel data for a specific location, the following steps can be followed. First, field measurements are conducted in the specific location to obtain downlink channel data. As the field data is utilized to capture the primary propagation characteristics of the channel, a limited amount of field data is typically sufficient. Second, the main channel parameters of the data are extracted, primarily including the MPC parameters and cluster parameters as mentioned in Section \ref{modeling}. Third, these parameters are utilized to analyze the statistical characteristics of the channel, effectively representing the wireless propagation characteristics. Subsequently, these statistical parameters are transmitted to the BS and used to update the parameters of the standard TR 38.901 channel model, resulting in the creation of the SSCM. Finally, the SSCM is employed to generate simulation channel data corresponding to the specific location. The aforementioned steps provide an overview of the proposed channel modeling aided data augmentation method, as illustrated in Fig. \ref{flowpath}.

To further enhance the precision and flexibility of SSCM on describing a specific place, the Poisson distribution is used to describe the number of clusters, which may bring an innovation to the typical TR 38.901. In typical TR 38.901, the cluster number is a constant value. It is generally regarded that the clusters are formed by the scatters in the wireless environment, some of which are steady clusters for example the buildings or trees, while some of which are dynamic or temporary clusters for example the cars or pedestrians. At different time, the cluster may change even in a fixed place. Therefore, it is better to describe the variability of the clusters in the environment, as the temporary clusters change dynamically in the actual systems. Motivated by the above requirements, the Poisson distribution is used to fit the number of clusters. Obviously, a varying number of clusters is more suitable for the actual deployed system.


\begin{figure*}[!t]
\centering
\includegraphics[width=7.6in]{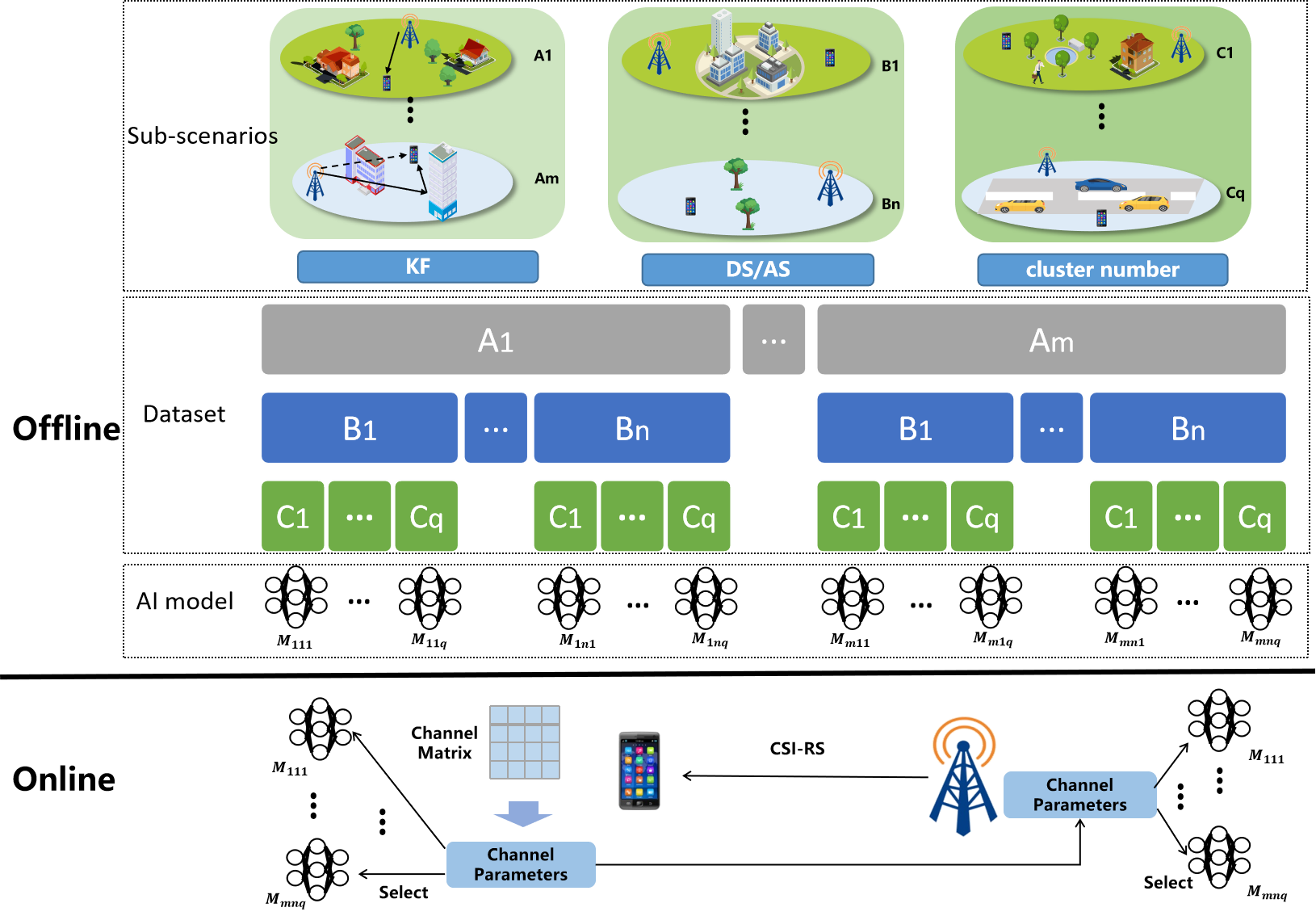}
\caption{Illustration for the dataset construction.}\label{dataset}
\end{figure*}
Our proposed data augmentation method inherently has the following benefits:
\begin{itemize}
\item Traditional field dataset relies on millions of field measurement to construct the dataset. While, our proposed strategy only needs to collect a limited number of field channel data, for example in Section \ref{field_dataset} 800 field samples are sufficient to construct the SSCM dataset. Therefore, the SSCM can immensely reduce the field measurement expense.
\item In our strategy, the air interface expense is only the stochastic characteristics parameters of the channel. While, in the field dataset strategy, it has to transmit the whole field dataset from the UE to the BS, which brings huge expenses to the air interface.
\item The SSCM inherits the property of the GBSM, that is the channel multipath parameters obey a certain distribution rather than a determinate direction, for example, the multipath angle of departure obeys inverse Gaussian distribution, multipath DS obeys logarithmic normal distribution and so on \cite{901}. This brings diverse channel characteristics to the simulation dataset, which no doubt benefits to the generalization of the autoencoder.
\item As the UE can obtain the channel statistics parameters from the field channel data, the channel statistics parameters can serve as an efficient tool for the AI model monitoring. Once the channel statistics parameters undergo a certain degree of change, which means the wireless propagation environment changes, then it reminds us to check if the current AI model still works efficiently.
\end{itemize}


\subsection{Dataset Construction}\label{dataconstru}

The CMDG serves to reduce data collection expenses and air interface overhead. However, it's important to note that the process of generating the dataset by the BS using the SSCM simulator, as well as the subsequent use of the produced dataset to train the AI model, are both time-consuming operations. In this section, a strategy for constructing the dataset is proposed to reduce the delay in dataset generation and the training delay of the AI model. The TR 38.901 typically describes a general scenario, which may not be detailed enough to generate a channel dataset for a specific scenario. This limitation prompts us to categorize the scenarios outlined in TR 38.901 into more detailed sub-scenarios based on the channel statistical parameters, such as AS, DS, KF, and others.

For example, AS or DS can help distinguish the complexity of the scattering environment, both in separate space and time domains, as a large AS/DS value typically corresponds to an environment with abundant scatters. Additionally, the KF can help distinguish between line-of-sight (LOS) or non-line-of-sight (NLOS) propagation scenarios, as a smaller KF indicates a higher probability of an NLOS scenario.


Based on extensive field measurements, we can develop a methodology to categorize the scenarios outlined in TR 38.901 into N sub-scenarios based on channel statistical parameters. Assuming the KF, AS/DS, and cluster number each divide the scenario into m, n, and q sub-scenarios, respectively. For a typical UMa scenario, it can be divided into $N = m * n * q$ sub-UMas. We can then construct N datasets corresponding to the sub-UMas using the approach mentioned in the data augmentation part, simultaneously obtaining $N$ sets of channel parameters corresponding to the $N$ sub-UMas.

When a UE enters a new UMa scenario, it can acquire the channel statistic parameters and upload them to the network. The network can then directly select one or several sets of sub-UMa datasets by comparing the uploaded channel parameters with the parameters of the $N$ sub-UMas. This approach helps to reduce the generation delay. Furthermore, we can pre-train the AI model for each sub-scenario in advance using the corresponding dataset. If a single sub-UMa dataset is chosen, the training delay can also be reduced.



\begin{figure*}[!t]
\centering
\subfloat[]{\includegraphics[width=3.3in,height=2.5in]{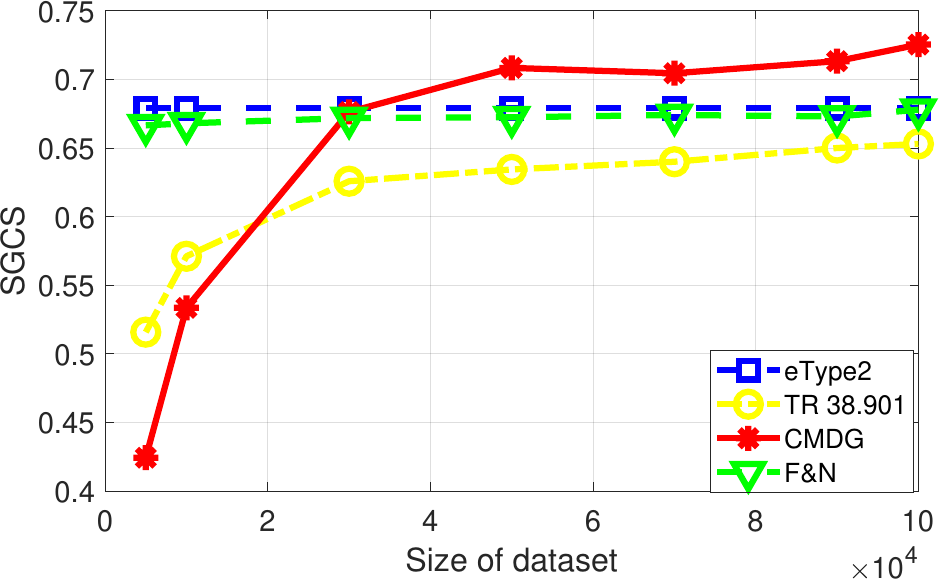}%
\label{fig_first_case}}
\hfil
\subfloat[]{\includegraphics[width=3.3in,height=2.5in]{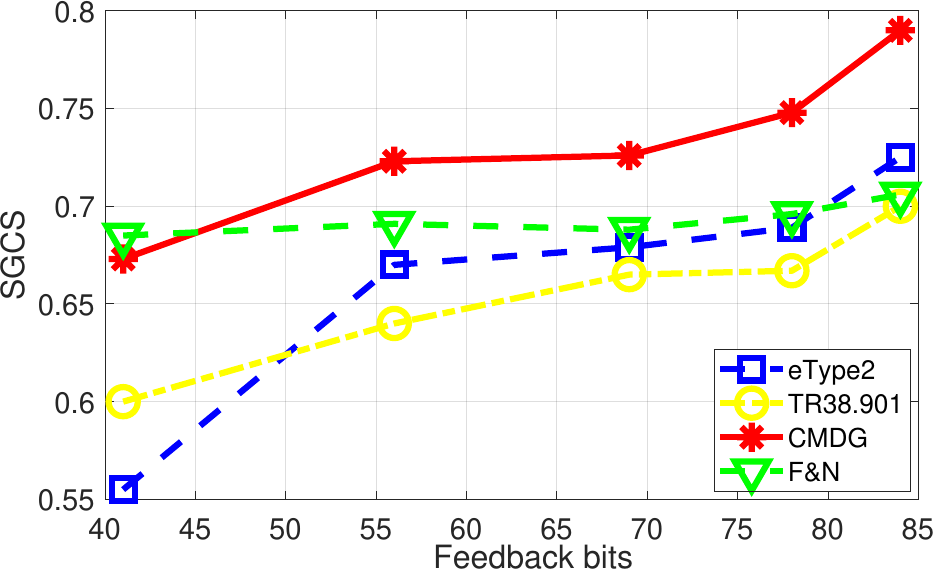}%
\label{fig_second_case}}
\caption{SGCS performance: a) performance comparison in different samples; b) performance comparison in different feedback bits.}
\label{fig4}
\end{figure*}

\section{Simulation Results and Discussions}
\subsection{Experiment Setting and Evaluation Metric}\label{field_dataset}
The field measurement data in a typical UMa scenario are used, where frequency channel matrix $H_{f}$ is provided. Please refer to \cite{dataset} for more detailed measurement configurations and data information. Three training datasets are constructed respectively with field channel dataset, TR 38.901 simulation dataset, and CMDG-based simulation dataset. Specifically, the field dataset is constructed with 5 sampling groups randomly selected in the AB path, in each group 5 sites are randomly selected and all the heights and directions are selected, totally 800 channel samples are selected. Similarly, the testing dataset is constructed randomly selected from the residual groups in the AB path, in each group 4 sites are randomly selected and all the heights and directions are selected, totally 640 channel samples are selected. The training set and the testing set are completely separated by the groups, whose distance is at least 15 meters. In this way, we can verify the generalization capability of the AI model. As the field channel data in a path is not large enough for the autoencoder being trained to converge, noise injection technique\cite{noise} is used to increase the dataset size. In the CMDG, the 800 field samples are used to accomplish our proposed data augmentation method.

The average squared generalized cosine similarity (SGCS)\cite{meta} is used to evaluate the CSI compress and recovery precision.

\subsection{Performance of CMDG}



In this section, we carry out simulations to evaluate the performance of CMDG. We compare the proposed algorithm with the following three benchmark schemes: 1) field channel dataset with noise injection (F\&N) based scheme. 2) TR 38.901 simulation dataset based scheme. 3) eType \uppercase\expandafter{\romannumeral2} codebook feedback scheme.


Figure \ref{fig4} (a) shows the SGCS performance versus the size of training dataset for different schemes. The feedback bits of the autoencoder is 56. As the F\&N uses the field channel data, it converges the fastest, and almost keep stable with the increase of dataset. This is because the noise injection cannot mine the intrinsic characteristics of the channel data, thus does not enrich sufficient diversity to the dataset. The CMDG outperforms eType \uppercase\expandafter{\romannumeral2} and F\&N at large size of dataset. And TR 38.901 exhibits the lowest SGCS. Figure \ref{fig4} (b) demonstrates the SGCS performance versus different feedback bits for different schemes. Sufficient large dataset is used to guarantee the convergence of all the schemes. It can be observed that, for all the schemes, the SGCS increases with the increase of feedback bits. The CMDG obtains the best SGCS performance. This is because the CMDG brings abundant channel characteristics corresponding to the actual environment, thus it possesses strong generalization capability in comparison with benchmarks.

\subsection{Performance of Dataset Construction}
In this section, four sets of simulation dataset are generated with TR 38.901 simulator to evaluate the performance of our proposed strategy of dataset construction. The carrier frequency is 2.6 GHz, the bandwidth is 20 MHz, the transmitter and receiver antennas are 8 and 4, respectively. The detailed channel parameters are illustrated in Table \ref{table} all in UMa scenario, where $lg$ represents the 10-base logarithm, $ASD$ is the azimuth angle spread of departure, $\mu$ and $\sigma$ represent the mean value and the standard deviation. From the table, we can see that the four datasets are generated with different channel statistic parameters, representing four kinds of sub-UMa scenario with different scattering environment.

Assuming the parameters and dataset of A, B, C are pre-measured and constructed in three different sub-UMa scenarios, and parameter set D corresponds to a measured channel in a newly deployed scenario. By comparison, we can see that parameter set B has the highest similarity with scenario D. From Section \ref{dataconstru}, we know that the dataset and AI model of B can be used to accomplish the CSI compression and feedback task in scenario D.

\begin{table}
\begin{center}
\caption{The channel parameters of the simulation dataset.}
\label{tab1}
\begin{tabular}{| c | c | c | c | c |}
\hline
Dataset & A & B & C & D\\
\hline
$\mu_{lgDS}$ & -7.6 & -6.8 & -6.0 & -6.6 \\
\hline
$\sigma_{lgDS}$ & 0.7 & 0.675 & 0.65 & 0.66\\
\hline
$\mu_{lgASD}$ & 1.26 & 0.7 & 1.6 & 0.75 \\
\hline
$\sigma_{lgASD}$ & 0.3 & 0.25 & 0.28 & 0.24 \\
\hline
$\mu_{KF}$ & 10 & 8 & 7 & 8.3 \\
\hline
$\sigma_{KF}$ & 4 & 3 & 4 & 2.8 \\
\hline
testing SGCS performance & 0.8655 & 0.9127 & 0.8775 & 0.9133 \\
\hline
\end{tabular}\label{table}
\end{center}
\end{table}

The four AI models are trained with the four training datasets constructed with parameter sets of A, B, C, and D, respectively. Then, the four AI models are tested with the test dataset constructed with D. Table \ref{table} illustrates the performance of the four AI models tested on dataset D. From the simulation results we can see that there is only small mismatch between the performances of B and D, that is only 0.066\% performance degradation. Simulation results verify that, through carefully design, the dataset construction method can achieve comparative performance with the CMDG. Besides, the SGCS of simulation dataset is higher than that of field dataset, as there is relative small distribution difference between the training dataset and the test dataset in the simulation dataset.


\section{Standardization Impact}
The minimization of drive tests (MDT) \cite{mdt} is utilized to monitor the state and performance of communication networks by gathering data from the UEs. It primarily consists of immediate MDT and logged MDT. Specifically, immediate MDT involves measurements by UEs in a connected state, reporting the results to the network periodically or when event-triggered. On the other hand, logged MDT implements measurement logging by UEs in an idle or inactive state, reporting to the network upon event-triggered occurrences, such as quantity-based events. Overall, MDT is a convenient and cost-effective method for data collection, as it does not require additional equipment.

In the current MDT specification, the criteria for large-scale channel quality are measured and reported, such as reference signal receiving power (RSRP), reference signal received quality (RSRQ), and others. However, for the proposed CMDG, the measurement and reporting of statistical parameters related to CSI, such as DS, AS, KF, cluster-related parameters, are not yet supported in MDT. Therefore, it is proposed to specify these statistical parameters in MDT and enhance the corresponding UE capabilities. Additionally, the logged MDT is currently only supported in idle or inactive states. However, the reference signals (RS) (i.e., CSI-RS) used to extract channel parameters are only transmitted in the connected state. Thus, the connected state configuration and corresponding measurements should be enhanced in the logged MDT.

\section{Conclusions and Future Work}

In this study, we analyzed the challenges faced by the AI-enabled autoencoder in CSI feedback, which include the overhead of dataset collection, weak generalization, and training strategy. To address these challenges and contribute to future standardization, we presented a channel modeling-enabled data augmentation method that comprehensively considers data collection, model generalization, and model monitoring. Based on the CMDG, we further proposed a strategy for dataset construction, pointing out a promising direction for dataset improvement. Simulation results demonstrated that CMDG significantly enhances SGCS performance compared to benchmarks. Additionally, we discussed the impact of standardization in our findings.

Research on the integration of AI and air interface is still in its early stages, and further tests are needed to verify the feasibility of our proposed dataset construction strategies. More field measurements should be conducted across various scenarios to establish a quantitative methodology for determining which channel statistics parameters to use and how many categories to divide based on each channel parameter. Furthermore, we highlighted the potential use of channel reciprocity to reduce the overhead of UE measurement and reporting. Despite only partial channel reciprocity between uplink and downlink channels in FDD, uplink channel measurement and statistical analyses may be sufficient to decide the sub-scenarios. If so, the sub-scenarios can be directly selected by the BS via uplink measurement, eliminating the need for additional downlink measurement and reporting by the UE.

\section*{Acknowledgments}
The authors would like to thank Tao Jiang, Xudong Wang and Hanning Wang, who have supported our work on this article.

%



%

\newpage

\end{document}